\documentclass[pdf]{iucr}
\usepackage{amsmath}
\usepackage{amssymb}
\usepackage{amsthm}
\usepackage{xcolor}
\usepackage{bm}
\usepackage{booktabs}
\usepackage[normalem]{ulem}
\usepackage{soul}  
\usepackage{footmisc}
\usepackage{hyperref}
\usepackage{lineno}
\usepackage{siunitx}
\usepackage{float}
\newcommand{\cd}[1]{\textcolor{teal}{CD: #1}}
\newcommand{\tr}[1]{\textcolor{orange}{TR: #1}}

\newcommand{\fs}[1]{\textcolor{blue}{FS: #1}}

\journalcode{S} 

\begin{document}


\title{Diamond compound refractive lenses for high energy Dark Field X-ray Microscopy}

\cauthor[a,b]{S.}{Staeck}{stest@dtu.dk}
\author{}{}    
\author[a]{C.}{Yildirim}

\author[c]{F.}{Seiboth}
\author[a]{T.}{Manning}
\author[a]{T.}{Roth}
\author[d]{J.C.}{Stinville}
\author[a]{C.}{Detlefs}

\aff[a]{European Synchrotron Radiation Facility, \city{Grenoble}, \country{France}}
\aff[b]{Technical University of Denmark, Department of Physics, \city{Kgs. Lyngby}, \country{Denmark}}
\aff[c]{Centre for X-ray and Nano Science CXNS, Deutsches Elektronen-Synchrotron DESY,\city{Hamburg}, \country{Germany}}
\aff[d]{Materials Science \& Engineering Department, University of Illinois Urbana-Champaign, \city{Urbana}, \country{United States}}

\date{\today}

\maketitle


\begin{abstract}
Compound-refractive lenses (CRL) are a type of x-ray optics that find widespread applications as focusing and imaging lenses. The choice of material is one of the most defining properties of these lenses. In this work, we present a CRL made out of diamond. It provides an advantageous balance between refractivity and absorption, along with good manufacturability. Compared to Be CRLs, it features a higher optical density and thus enables application at higher photon energies without relying on large lens stacks or very small radii of curvature, which are challenging to manufacture. A diamond CRL is characterized for use as an objective for Dark-field X-ray Microscopy (DFXM) at the ID03 beamline of the European Synchrotron Radiation Facility (ESRF) and compared to Al and Be CRLs at \SI{17}{\keV}, \SI{33}{\keV} and \SI{37}{\keV}. Increasing the photon energy in DFXM from \SI{17}{\keV} to \SI{37}{\keV} opens up the possibility to investigate new sample systems, that were previously opaque to low energy x-ray radiation. The capability of the diamond CRL at \SI{33}{\keV} is illustrated through DFXM measurements on two \SI{0.5}{\milli\metre}-thick iron-based samples, which cannot be probed at \SI{17}{\keV}.

\end{abstract}


\section{Introduction}

The concept of using Compound Refractive Lenses (CRLs) to focus x-rays was first introduced in a 1994 patent by Tomie. 
Tomie proposed that, despite the extremely low refractive indices of materials at x-ray wavelengths, effective focusing could be achieved through a series of concave lenses fabricated from low atomic number (low-$Z$) materials. This transmissive lens-based approach stood in contrast to established methods relying on grazing incidence reflection or diffractive optics such as Fresnel zone plates.

In 1996 Snigirev \textit{et al.} demonstrated a practical implementation of CRLs at a synchrotron radiation source \cite{Snigirev1996}. 
By drilling cylindrical holes into an aluminum block, they created the first working prototype of a refractive x-ray lens, thereby establishing the feasibility of compact, in-line x-ray focusing optics for high-energy applications. 
Since then, CRLs have become a widely adopted solution in synchrotron and XFEL beamlines, used in diverse applications such as phase-contrast imaging, tomography, full-field microscopy, and nano-focused x-ray diffraction \cite{Lengeler1999}.

Manufacturing advances have greatly improved the optical quality of CRLs, enabling sub-micrometer precision and reduced aberrations. Techniques such as deformation via indentation, deep x-ray lithography, LIGA processing, laser micromachining, and 3D nanoprinting now allow the fabrication of lenses from a broad range of materials \cite{Roth2017}. These developments have expanded the design space to include transfocators for focal length adjustment \cite{Snigirev2009a,Vaughan2011}, beam-shaping geometries like alligator lenses \cite{Cederstrom2000}, and CRLs made from advanced polymers such as SU-8 \cite{krywka2016polymer}.

\begin{figure}
    \centering
    \includegraphics[width=0.9\columnwidth]{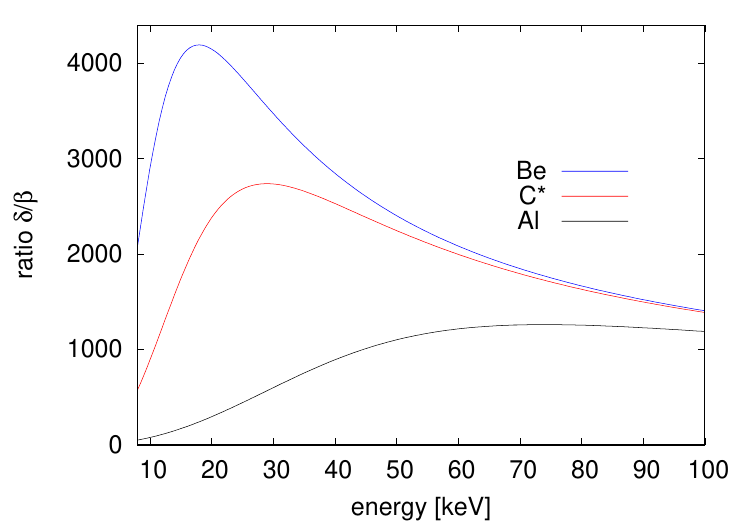}
    \caption{Ratio of $\frac{\delta}{\beta}$ for the three most commonly used lens materials.}
    \label{fig:deltaoverbetar}
\end{figure}

Material selection is a key factor in compound refractive lens (CRL) design, as both focusing efficiency and x-ray transmission are determined by the complex refractive index $n = 1 - \delta + i\beta$. The real part, \(\delta\), governs the lens' focusing power, while the imaginary part, \(\beta\), describes absorption. For efficient lens systems, especially those with many elements, it is crucial to maximize \(\delta\) while minimizing \(\beta\) to preserve beam intensity. Fig.\ref{fig:deltaoverbetar} shows the ratio $\frac{\delta}{\beta}$ for Al, Be and diamond. Among available materials, beryllium (Be) has long been considered ideal for hard x-ray applications due to its favorable combination of high transmission and sufficient focusing strength~\cite{serebrennikov2016}. Be CRLs have been successfully deployed in both condenser and objective roles, including at the ESRF’s ID06-HXM beamline for dark-field x-ray microscopy \citeaffixed{Kutsal2019}{DFXM, }{}, now upgraded as ID03~\cite{Isern2025}, and in x-ray free-electron laser (XFEL) facilities~\cite{heimann2016, Zozulya2019CRL, Dresselhaus2023}.

Despite these advantages, the practical use of Be is increasingly constrained. The material is toxic, brittle, and expensive, with growing difficulties in procurement. Moreover, structural imperfections such as voids, inclusions, and grain boundaries can introduce small angle scattering, degrading coherence and image quality~\cite{Timmann2009, Roth2014, Lyatun2020}. These issues have led to the discontinuation of commercial Be lens production, for instance, by RXOPTICS, accelerating the search for alternative materials better suited to modern synchrotron and XFEL environments.

Aluminum (Al) lenses are relatively cheap and easy to manufacture and therefore display good imaging performance with small lens errors \cite{snigirev1998focusing,schroer2000compound}. They are also relatively stable under thermal load in high-flux applications. However, diamond and beryllium offer better transmission performance, especially at lower energies (below \SIrange[]{20}{30}{keV}).

Nickel (Ni) CRLs offer a robust option for focusing high-energy x-rays \cite{nazmov2005liga,Andrejczuk2014,brancewicz2016high}. Since Ni is a rather heavy element, it has a high diffractive power combined with high absorption losses. Therefore, it is mainly applicable for high energy applications ($>  \SI{30}{keV}$), where the absorption losses are tolerable. 

Silicon (Si) is a radiation-hard material with excellent thermal and mechanical stability, increasingly used for CRLs in hard x-ray optics \cite{Schroer2003,Schroer2005,simons2016full,ruett2022etched}. Again, compared to Be or diamond lenses, the diffraction to absorption ratio is smaller, especially at lower energies.

Polymeric CRLs fabricated from SU-8 negative photoresist offer a practical solution for x-ray focusing in the \SIrange{5}{30}{keV} range \cite{marschall2014x,Simons2015, yildirim2020probing, hlushko2020dark}. They are comparatively easy and reproducible to manufacture and feature low absorption, but lack radiation hardness and thermal stability.

Among all candidates, diamond has emerged as the most likely material for next-generation CRLs. It combines a relatively low $\beta$ (high transmission) with high $\delta$ (strong focusing) in the hard x-ray regime. 
Most importantly, its superior thermal conductivity makes diamond lenses uniquely suited for high-heat-load applications, such as white and pink beam transfocators or XFEL optics \cite{Antipov2016b, Isern2025,Yildirim2025, Labella2025, Dresselhaus2023, Irvine2025}. The intrinsic radiation hardness and structural integrity of diamond allow stable performance under extreme flux conditions. The single-crystal substrates used for manufacturing diamond CRLs have excellent homogeneity, therefore exhibit much lower small-angle scattering than polycrystalline Be, especially when the optical surfaces are polished \cite{Timmann2009,Gasilov2017,Celestre2022}.
Despite fabrication challenges, requiring focused ion beam (FIB) milling or laser micromachining, recent advances have demonstrated reliable production of diamond CRLs with excellent optical quality \cite{Snigirev2002, Antipov2016, Polikarpov2016, Seiboth2017, Celestre2022,Wang2025}.


In this work, we compare and quantify the performance of compound refractive lenses (CRLs) fabricated from single-crystal diamond against conventional beryllium and aluminum counterparts under both bright-field and dark-field imaging conditions. Using a systematic set of test measurements, we assess resolution, field-of-view and transmission efficiency across a range of photon energies from \SIrange[]{17}{37}{keV}. Diamond CRLs were benchmarked against Be CRLs at \SI{17}{keV} and Al CRLs at \SI{33}{keV}, using standardized JIMA resolution targets and Si checkerboards to characterize optical quality and aberrations. Additionally, DFXM measurements at \SI{33}{keV} were performed on recrystallized ferritic iron and Invar alloy samples to evaluate the suitability of diamond CRLs for high-resolution diffraction contrast imaging, including quantitative strain and mosaicity mapping. This comprehensive dataset provides a practical basis for selecting CRL materials for high-flux synchrotron environments and emerging fourth-generation light sources. The following sections describe the experimental configurations, imaging modalities, and evaluation metrics used in this comparative study.


\section{Experimental setup}


Different types of CRLs were characterized regarding their resolving power at different energies, the extent of the spatial distortion and their field-of-view (FOV). Furthermore, two application samples were investigated to showcase the potential of the diamond CRLs for DFXM in section \ref{sec:microscopy}. All these measurements were performed at the beamline ID03 of the European Synchrotron Radiation Facility (ESRF), which is described in the following subsection. The characterization results of the different CRLs are presented below.

\subsection{Lens stacks} 
Four CRL stacks have been tested in bright field x-ray microscopy: 
\begin{itemize}
\item At \SI{17}{\keV}, an $N=20$, $R=\SI{25}{\micro \metre}$ diamond lens stack and an equivalent $N=88$, $R=\SI{50}{\micro \metre}$ Be lens stack. 
\item At \SI{33}{\keV}, an $N=124$, $R=\SI{30}{\micro \metre}$ Al lens stack and an equivalent $N=70$, $R=\SI{25}{\micro \metre}$ diamond lens stack. 
\item The latter were also used for dark field X-ray microscopy, both at \SI{33}{\keV} and at \SI{37}{\keV}. 
\end{itemize}
Table \ref{Table:lensStacks} summarizes the geometrical dimensions of the lens stacks. 

\begin{table}
\setlength{\tabcolsep}{6pt}
\caption{\label{Table:lensStacks}Summary of used lens stacks (CRLs)}
\begin{tabular}{cl|crcrcc}
\toprule
E     & name & \footnotesize material & $N$ & $R$    &$l_\textrm{lens}$& $A_\textrm{phys}$ & $L_\textrm{stack}$ 
\\
\footnotesize$[\textrm{keV}]$ & & & &\footnotesize[\SI{}{\micro \metre}]& \footnotesize[\SI{}{\micro \metre}] & \footnotesize[\SI{}{\micro \metre}] & \footnotesize[mm] 
\\
\midrule
17 & $\mathrm{Be}_{88}^{\SI{50}{\micro \metre}}$ & Be & 88 & 50 & 1000 & 440 & 148.8 
\\
 & $\mathrm{C}_{20}^{*\SI{25}{\micro \metre}}$ & \footnotesize Diamond & 20 & 25 & 250 & 120 & 12
 \\
\midrule
33 &$\mathrm{C}_{70}^{*\SI{25}{\micro \metre}}$ & \footnotesize Diamond & 70 & 25 & 250 & 120 & 42
\\
 & $\mathrm{Al}_{124}^{\SI{30}{\micro \metre}}$ & Al & 124 & 30 & 600 & 260 & 148.8
 \\
 \bottomrule
\end{tabular}
\end{table}

The diamond lens stacks were produced at DESY using femtosecond laser ablation machining into single-crystalline chemical vapor deposited (scCVD) diamond plates \cite{Wang2025} on either side, leading to bi-concave lenses. No polishing step was applied after the laser-ablation. The surface micro-roughness after the ablation is typically below \SI{200}{nm} (RMS). As transmission through diamond is high, especially at \SI{33}{\keV}, a tungsten pinhole of \SI{120}{\micro \metre} diameter, slightly smaller than the diamond lens physical aperture, was used to reduce unwanted background from x-rays passing by the side of the physical lens aperture. After each ten lens plates, a previously determined correction optics phase plate was added to the stack in order to compensate for fabrication errors \cite{Seiboth2017}. For the high energy lens stack, this meant seven correction plates. The benefit of distributing many correction plates along the lens stack is that such a corrected lens stack can be used over a wide x-ray energy range without deteriorating the quality of the correction \cite{WDB+25}.

The Be lens stack was assembled using Be lenses purchased from RXOPTICS, with a mixture of Be grades coming from powder metallurgy and others from rolled ingot grades \cite{Dombrowski1997,Roth2014}. The latter show less small angle x-ray scattering (SAXS) background, but due to the high number of required lenses, we were forced to use a mixture, as currently, no lenses with rolled ingot grades are commercialized. 

The Al lens stack was produced at the ESRF using 99.999\% pure Aluminium discs of \SI{4}{mm} diameter and \SI{1}{mm} thickness, polished to \SI{40}{nm} rms surface roughness and pre-thinned to \SI{600}{\micro \metre}, using flat punches. As with the Be lenses, the parabolic lens shape is obtained via a deformation process using two punches, in our case with parabolic $R=\SI{30}{\micro \metre}$ apex-radius tips, made in hardened steel. The width at the apex was \SI{15}{\micro \metre} on average.

The full-field imaging comparison of the different CRL stacks was carried out at the ID03 beamline at ESRF.

\subsection{Beamline setup}
\begin{figure}
    \begin{center}
    \includegraphics[width=1\linewidth]{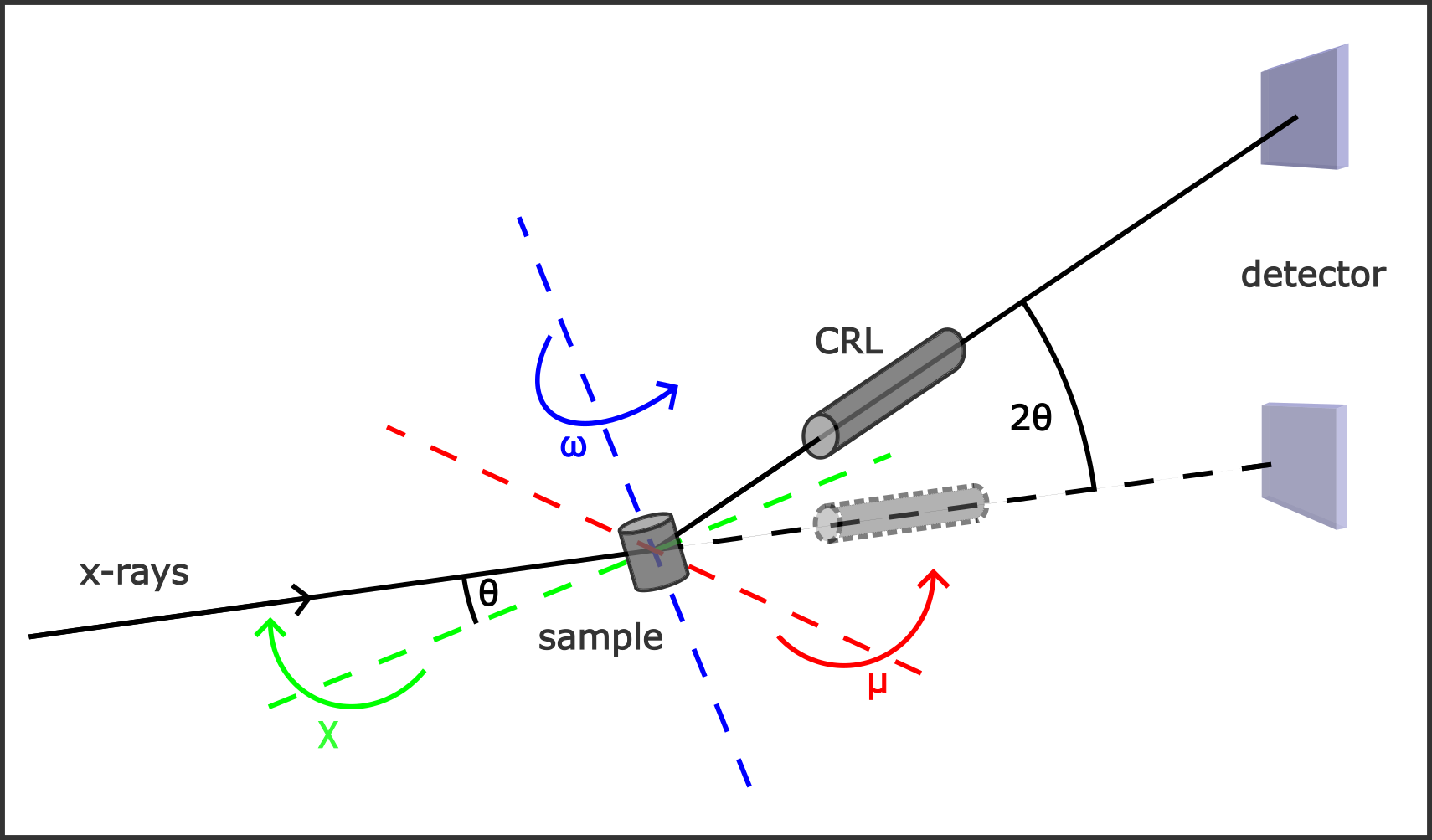}
    \caption{Schematic of the DFXM setup at the ID03 beamline at the ESRF. Depicted is the basic setup geometry with the dark-field (solid line) and bright-field (dashed line) configuration. In dark-field mode, the incoming beam is diffracted at the sample and the diffracting part of the sample is imaged by the CRL on the detector. Here, $\theta$ describes the Bragg angle. The CRL objective and the detector are positioned at an angle of $2 \theta$ in respect to the optical axis. $\mu$, $\chi$ and $\omega$ are depicted as well. $\mu$ and $\chi$ can be used to investigate the mosaicity of a sample by bringing misoriented parts of the sample into diffraction condition. In bright-field mode, the CRL and the detector are in line with the incoming x-ray beam and the sample is imaged via absorption and phase contrast.}
    \label{beamline}
    \end{center}
\end{figure}

The ID03 beamline at ESRF \cite{Isern2025} specializes in dark-field x-ray microscopy (DFXM). This technique utilizes x-ray optics in the Bragg-diffracted beam to image the crystalline microstructure of samples. Using the diffraction contrast, the technique is sensitive to the orientation (mosaicity) and $d$-spacing (strain) of lattice planes in the sample \cite{Simons2015,Poulsen2017}. A schematic of DFXM at the ID03 beamline is shown in Fig.~\ref{beamline}.

The beamline utilizes a multilayer and a crystal monochromator to monochromatize the x-ray beam generated by a cryogenic permanent magnet in-vacuum undulator (CPMU). The beam is shaped by several sets of slits before it hits the sample. ID03 has an energy range from \SIrange{11}{60}{\keV}; DFXM experiments are usually performed at \SI{17}{\keV} or \SI{19}{\keV}. A 1D Be CRL condenser may be used to generate a sheet beam for layered measurements of the sample (section topography). 
The sample is mounted on a hexapod, which is again mounted on a goniometer for sample alignment and measurement. The goniometer features two axes of translation and two axes of rotation, $\mu$ and $\omega$, see Fig.~\ref{beamline}. 
The magnified image is recorded by a ``far field detector'' at a distance off \SIrange{2.5}{5}{\metre} downstream of the sample. This is an indirect detector comprising a crystal scintillator, magnifying optics ($\times 2$ or $\times 10$), and an sCMOS camera (pco.Edge 4.2BI).

The goniometer angles $\mu$ and $\chi$ can be used to investigate the mosaicity of a sample to bring different, slightly misoriented sample parts into diffraction condition. The strain of a sample can be investigated by either scanning the scattering angle $2\theta$ by moving the objective and far field detector, or by scanning the photon energy. Such scans are typically combined with rocking scans ($\mu$).
Further experimental details can be found in \citeasnoun{Isern2025}.

In the measurements presented here, the sample to detector distance is \SI{5}{\metre}. The values for magnification, effective pixel size and exposure time are given in the respective subsections. In general, the magnification is around $M = 140-150$, including the magnification of the $\times 10$ far-field objective, and the effective pixel is typically around \SIrange{40}{50}{\nano\metre}. The exposure time was either \SI{1}{\second} or \SI{2}{\second}, depending on the measurement.


\section{Results and Discussion}

\subsection{Bright field X-ray microscopy}
This section presents the resolution, radial distortion, and field of view in bright field ($2\theta = 0$).
\subsubsection{Resolution}
\begin{figure}
    \begin{center}
    \includegraphics[width=1\linewidth]{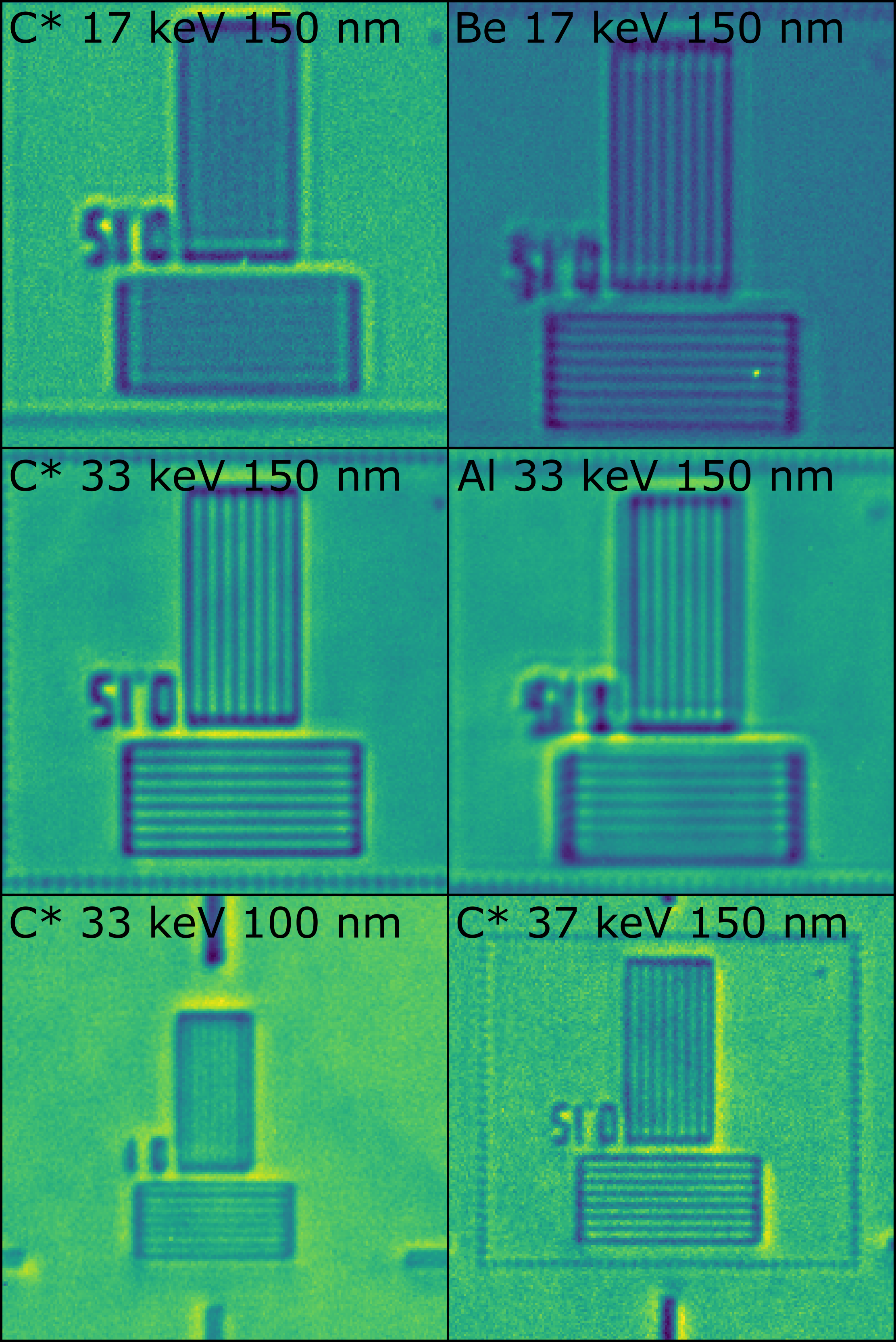}
    \caption{JIMA test pattern structures for different combination of CRL objectives, x-ray energies and structure sizes.}
    \label{resolution}
    \end{center}
\end{figure}

The resolving power of the diamond CRL was tested at different energies using a JIMA test pattern \cite{JIMA100}. The results are compared to those of the Be CRL currently in use at low energies (\SI{17}{\keV}) and to an Al CRL at higher energies (\SI{33}{\keV}). Furthermore, test patterns have been recorded with the diamond CRL at \SI{37}{\keV} to investigate the potential to use this lens at even higher energies. The diamond lens stack has been adjusted according to the energy as presented in Table \ref{Table:lensStacks}. For the lower energy experiment, the energy in case of the diamond CRL was slightly altered to \SI{17.7}{\keV}, to keep a similar focal distance and magnification as the Be CRL. We will refer to this energy as \SI{17}{\keV} from here on as well, for the sake of simplicity. For the measurements, the \SI{100}{nm}, \SI{150}{nm} and \SI{200}{nm} structure size test patterns on the JIMA resolution target were imaged in bright field with the pco far-field detector and the $10\times$ objective. From the images the Michelson contrast was calculated in the vertical and horizontal direction. The results are presented in Table \ref{Table:resolution}, together with the maximum contrast measured in the images on a large structure on the JIMA resolution target. A choice of images of the JIMA test structure for different combinations of CRLs, photon energies and structure sizes can be seen in Fig. \ref{resolution}. The values for magnification and effective pixel size for the far-field detector with the $10\times$ objective and exposure time of the characterisation measurements are given in Table~\ref{Table:params}. The magnification includes the magnification from the objective. For the $2\times$ objective, the values for magnification and effective pixel size shown in Table~\ref{Table:params} scale with a factor of 5, accordingly. 

\begin{table}
\setlength{\tabcolsep}{6pt}
\caption{\label{Table:resolution}Horizontal and vertical Michelson contrast for the different CRLs evaluated and the maximum Michelson contrast measured for each energy.}
\begin{center}
\begin{tabular}{c|c|crcrcc}
\toprule
E     & size& name & vertical  & horizontal & maximum &
\\
{} [\SI{}{keV}]&  [\SI{}{\micro \metre}] & &contrast &contrast & contrast 
\\
\midrule
17 & 100 &  $\mathrm{Be}_{88}^{\SI{50}{\micro \metre}}$  & \num{0.015 \pm 0.003} &\num{0.009 \pm 0.004}  & \num{0.142 \pm 0.003}
\\

 & 150 &  $\mathrm{Be}_{88}^{\SI{50}{\micro \metre}}$  & \num{0.0561 \pm 0.0018} & \num{0.068 \pm 0.003} & 
\\
 && $\mathrm{C}_{20}^{*\SI{25}{\micro \metre}}$  &\num{0.0082 \pm 0.0008}  &\num{0.0078 \pm 0.0008}  & 
 \\
 & 200 &  $\mathrm{Be}_{88}^{\SI{50}{\micro \metre}}$  & \num{0.133 \pm 0.004} & \num{0.084 \pm 0.002} & 
\\
 && $\mathrm{C}_{20}^{*\SI{25}{\micro \metre}}$  &\num{0.096 \pm 0.002}  & \num{0.071 \pm 0.002}  & 
 \\
\midrule
33 &100&$\mathrm{C}_{70}^{*\SI{25}{\micro \metre}}$ & \num{0.006 \pm 0.001}  & \num{0.0034 \pm 0.0004} &  \num{0.042 \pm 0.006}
\\
 & 150 &$\mathrm{C}_{70}^{*\SI{25}{\micro \metre}}$ & \num{0.0331 \pm 0.0013}  & \num{0.0267 \pm 0.0019}& 
\\
 & &$\mathrm{Al}_{124}^{\SI{30}{\micro \metre}}$  & \num{0.0158 \pm 0.0013} &   \num{0.0249 \pm 0.0013} & 
 \\
  & 200 &$\mathrm{C}_{70}^{*\SI{25}{\micro \metre}}$ & \num{0.063 \pm 0.0008}  & \num{0.0647 \pm 0.0006} & 
\\
 & &$\mathrm{Al}_{124}^{\SI{30}{\micro \metre}}$  & \num{0.0448 \pm 0.0015} &   \num{0.0435 \pm 0.0013} & 
 \\
\midrule
37 &150&$\mathrm{C}_{70}^{*\SI{25}{\micro \metre}}$ &  \num{0.0163 \pm 0.0007}  &\num{0.0094 \pm 0.0003} &  \num{0.036 \pm 0.003}
\\
\bottomrule
\end{tabular}
\end{center}

\end{table}

\begin{table}
\setlength{\tabcolsep}{6pt}
\caption{\label{Table:params}Summary of the measurement parameters.}
\begin{tabular}{cl|crcr}
\toprule
E     & name &  $M$ & ef. pixel size   
\\
\footnotesize$[\textrm{keV}]$ & & & \footnotesize[nm] 
\\
\midrule
17 & $\mathrm{Be}_{88}^{\SI{50}{\micro \metre}}$ & 151.7  & 42.9  
\\
& $\mathrm{C}_{20}^{*\SI{25}{\micro \metre}}$ & 155.9 & 41.7 
\\
\midrule
33 &$\mathrm{C}_{70}^{*\SI{25}{\micro \metre}}$ & 140.8 & 46.2
\\
& $\mathrm{Al}_{124}^{\SI{30}{\micro \metre}}$ & 140.8 & 46.2 
\\

\midrule
37 &$\mathrm{C}_{70}^{*\SI{25}{\micro \metre}}$ & 112.7 & 57.7 
\\
\bottomrule
\end{tabular}
\end{table}

At \SI{33}{\keV}, the diamond CRL achieves better contrast and resolution than 
the Al lenses, while at \SI{17}{\keV} the Be lenses provide superior resolution 
to the diamond CRL. This ordering follows directly from the numerical apertures 
of the three lens systems, with $\text{NA}_\text{Be} > \text{NA}_{\text{C}^*} 
> \text{NA}_\text{Al}$. At \SI{17}{\keV}, the Be CRL reaches 
$\text{NA} = 5.47\times 10^{-4}$, more than twice that of the diamond CRL 
($\text{NA} = 2.25\times 10^{-4}$), which accounts for both the resolution 
and the contrast difference observed at this energy. For the Be and Al lenses 
the effective aperture $D_\text{eff}$ is set by absorption, whereas for the 
diamond CRL it is limited instead by the \SI{120}{\micro\metre} pinhole placed 
in front of the lens stack. At \SI{33}{\keV}, the diamond CRL reaches 
$\text{NA} = 2.02\times 10^{-4}$, exceeding that of the Al lenses 
($\text{NA} \approx 1.5\times 10^{-4}$) and combined with the shorter 
wavelength explains the gain in both contrast and resolution. Consistently, 
the diamond CRL at \SI{33}{\keV} resolves \SI{100}{\nano\metre} structures 
with a quality comparable to that achieved at \SI{150}{\nano\metre} at 
\SI{17}{\keV}. Furthermore, as shown in Fig.~\ref{resolution} (bottom right), 
the diamond CRL is capable of imaging \SI{150}{\nano\metre} structures even 
at \SI{37}{\keV}.

\subsubsection{Radial distortion and field-of-view}
In an effort to characterize radial distortions (pin-cushion or barrel) of the diamond and the Al CRLs at \SI{33}{\keV} and the diamond and the Be CRLs at \SI{17}{\keV} a \SI{10}{\micro \metre} checkerboard pattern was imaged in direct beam geometry. With this, the distortion of the straight lines of the pattern could be observed. The radial distortion is quantified by applying the first order of the Brown–Conrady radial distortion model \cite{Brown1966,Conrady1919}:

\begin{equation}
r_d = r_u \left( 1 + k_1 r_u^2\right)
\end{equation}

Here, $r_d$ is the distorted distance from the pixel in question to the distortion center (which is assumed to be the center of the intensity distribution), $r_u$ is the undistorted distance and $k_1$ is the parameter describing the direction and magnitude of the radial distortion. For $k_1 < 0$, $r_d$ is smaller than $r_u$ and the distorted pixel are shifted towards the distortion center. In this case, the image experiences a barrel distortion. Likewise, for $k_1 > 0$ the image experiences a pin-cushion distortion.\\
The parameter $k_1$ was determined by calculating a histogram of the marked edge points in dependence on the radial distance to the center of the intensity distribution. For this, all edges of the checkerboard pattern are marked by an algorithm, or in case this was not feasible, by hand. The center of mass of the intensity distribution was determined in the x- and y-direction of the image. The number of edge points $N \left(r \right)$ is proportional to the number radial distance $r$, because the area of the annulus $A \left(r \right)$ is proportional to the radial distance as well: 

\begin{equation}
A \left(r \right) = 2 \pi r \Delta r \Rightarrow N \left( r \right) \propto r
\end{equation}

Any radial distortion will influence the otherwise constant spacing between the edge points and their density and thereby $N \left(r \right)$. This will lead to a deviation of the linear relationship between $N \left(r \right)$ and $r$, which can be described with the Brown–Conrady radial distortion model. Specifically, the number of edge points is proportional to the following formula:

\begin{equation}
N \left(r \right) \propto r(1 + 4 k_1 r^2)
\end{equation}

$N \left(r \right)$ is then fitted with the following formula:

\begin{equation}
N \left(r \right) = a r(1 + b r^2)
\end{equation}

\begin{figure}
    \begin{center}
    \includegraphics[width=0.75\linewidth]{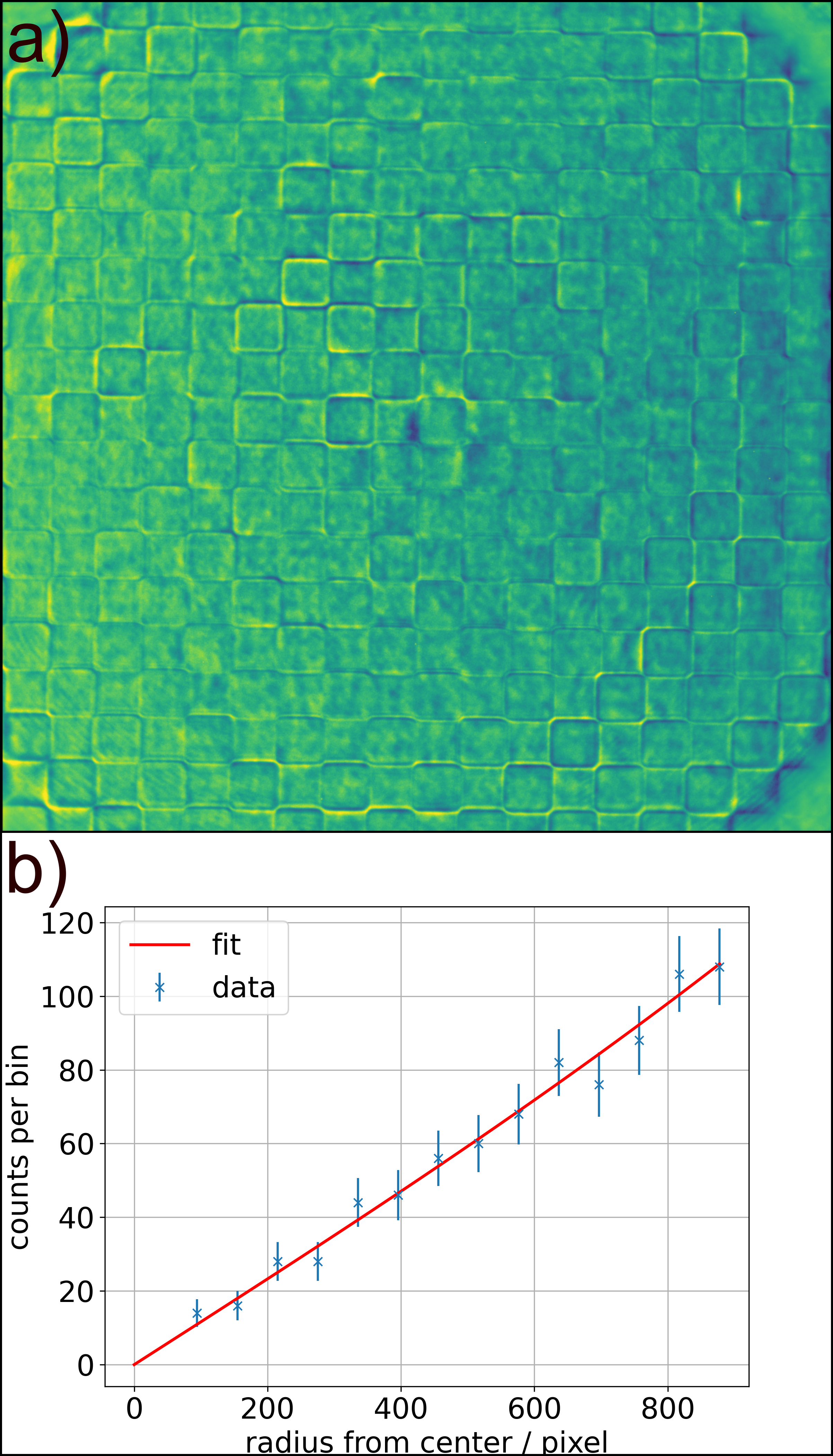}
    \caption{a) Checkerboard pattern imaged with the diamond CRL at \SI{33}{\keV}. b) Fit of the edge points per histogram bin in dependence on the radius from the center of mass of the intensity distribution of a). The error bars represent the statistical uncertainties. A slight deviation from the linear relation is visible.}
    \label{radial_distortion}
    \end{center}
\end{figure}

Here, $a$ is a proportionality factor depending on the edge point spacing and $k_1 = b/4$.\\
A cell close to that center was chosen as the center of the checkerboard pattern and used to measure the size of that cell in pixel, assuming a potential pin-cushion or barrel distortion to be close to the center of the intensity distribution. The checkerboard pattern is used as a scale for the calibration of the pixel size. In addition to that, the measurement of the checkerboard pattern is used to  estimate the field-of-view (FOV) of the different objectives at different energies by calculating the full-width at half-maximum (FWHM) of the intensity distribution. The effective size is given as well. The FOV values are averaged over the x- and y-direction of the image. Note that the measurements for \SI{33}{\keV} were performed with the $10\times$ objective on the pco far-field camera in comparison to the $2\times$ objective for the measurements at \SI{17}{\keV}. The $k_1$ values are given in \unit{\micro \metre^{-2}} for comparability. An examplary checkerboard image recorded with the diamond CRL at \SI{33}{\keV} and the fit of the histogram of edge points are shown in Fig. \ref{radial_distortion}.  The results are shown in Table \ref{Table:distortion}. \\

\begin{table}
\setlength{\tabcolsep}{6pt}
\caption{\label{Table:distortion}Radial distortion parameters $k_1$ for the different CRLs at different energies and their field-of-views (FOV).}
\begin{center}
\begin{tabular}{c|crcrcc}
\toprule
E     & name  & $k_1$ & FOV & ef. pixel size 
\\
\footnotesize$[\textrm{keV}]$ & &\footnotesize[\unit{ \micro \metre^{-2}}] &\footnotesize[\unit{\micro \metre}] & [\unit{\nano\metre}]
\\
\midrule
17 & $\mathrm{Be}_{88}^{\SI{50}{\micro \metre}}$ &  \num{1.3 \pm 0.4d-5} & 125.11 & 212.8
\\
 & $\mathrm{C}_{20}^{*\SI{25}{\micro \metre}}$ &\num{1.0 \pm 0.6d-5}  & 91.36  & 227.2
 \\
\midrule
33 &$\mathrm{C}_{70}^{*\SI{25}{\micro \metre}}$ &  \num{1.0 \pm 1.2d-5}  & 89.44 & 45.0
\\
 & $\mathrm{Al}_{124}^{\SI{30}{\micro \metre}}$ &  \num{0.3 \pm 1.2d-5} &   64.55 & 45.2
 \\
\bottomrule
\end{tabular}
\end{center}

\end{table}

The results show small values for $k_1$. Pin-cushion distortions are noticeable for all lenses, although negative $k_1$ are within the uncertainty budgets for the lenses at \SI{33}{\keV}. The uncertainties are quite high, which shows that the measurement method might not be suitable for quantifying radial distortions that small. For example, $k_1 = \num{1.3d-5}$\unit{ \micro \metre^{-2}} for the Be lenses at \SI{17}{\keV} induces a shift of about \SI{960}{\nano \metre} at a radius of \SI{42}{\micro \metre}. For most applications, the radial distortions for all lenses measured are negligible.
The FOVs differ rather drastically for different lenses, with the Be lenses exhibiting the largest FOV with \SI{125.11}{\micro \metre}. The diamond lenses have a FOV of around \SI{90}{\micro \metre} at both \SI{17}{keV} as well as \SI{33}{keV}, while the Al lenses have the smallest FOV with \SI{64.55}{\micro \metre}.\\

\subsection{High energy dark field X-ray microscopy}\label{sec:microscopy}


High-energy DFXM measurements above \SI{30}{\keV} have previously been reported only for a few cases using SU-8 CRL objectives at the ESRF ID06-HXM beamline, the former home of the Hard X-ray Microscope at ESRF \cite{Kutsal2019,hlushko2020dark, yildirim20224d}. Building on those demonstrations, the new diamond CRL objective was utilized here in two experiments showcasing the potential of diamond optics for high-energy DFXM. At \SI{33}{\keV}, two different iron-based samples were investigated. Operation at \SI{33}{\keV} within the geometrical constraints of ID03 enables access to sample systems that are strongly attenuating at \SI{19}{\keV} and below, thereby extending DFXM to thicker specimens and to compositions containing heavier elements. The results are presented in the following subsections.

\subsubsection{Recrystallized iron}
\begin{figure}
    \begin{center}
    \includegraphics[width=0.9\linewidth]{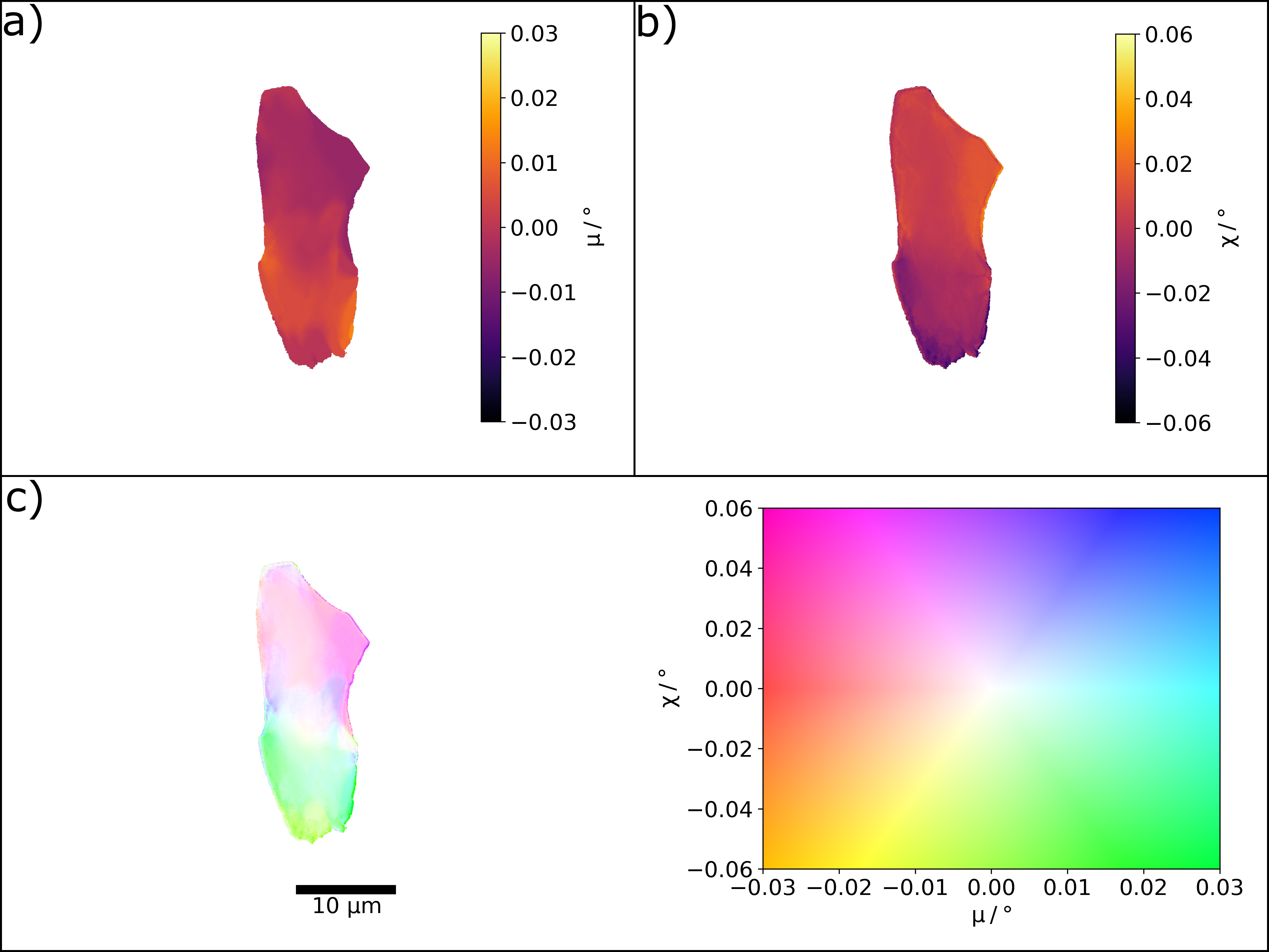}
    \caption{Grain~1 of the recrystallized iron sample imaged with the diamond CRL objective at \SI{33}{\keV}. 
a) Rocking angle $\mu$ at maximum intensity for each pixel. 
b) Rolling angle $\chi$ at maximum intensity for each pixel. 
c) Combined intragranular misorientation map referenced to the centre of the $\mu$ and $\chi$ rocking curves, revealing local lattice rotations associated with intragranular defects and elastic strain heterogeneities. 
The colormap and scale bar are indicated.}
    \label{rex_g1}
    \end{center}
\end{figure}

The iron sample investigated here was fully recrystallized high-purity iron and is similar to the material studied in \cite{Shukla2025,sanna2026}, where full details of the sample preparation and characterization are provided. Briefly, the material was cold rolled to a $90\,\%$ thickness reduction and subsequently annealed at \SI{700}{\celsius} for \SI{30}{\minute} to achieve a fully recrystallized microstructure. A needle-shaped specimen was extracted from the rolled plate by electrical discharge machining and electro-polished to remove surface damage. DFXM measurements were performed near the top section of the sample within a local cylindrical volume of approximately \SI{500}{\micro \metre} in diameter, where the average grain size was \SIrange[]{15}{20}{\micro \metre}. 

\begin{figure}
    \begin{center}
    \includegraphics[width=0.9\linewidth]{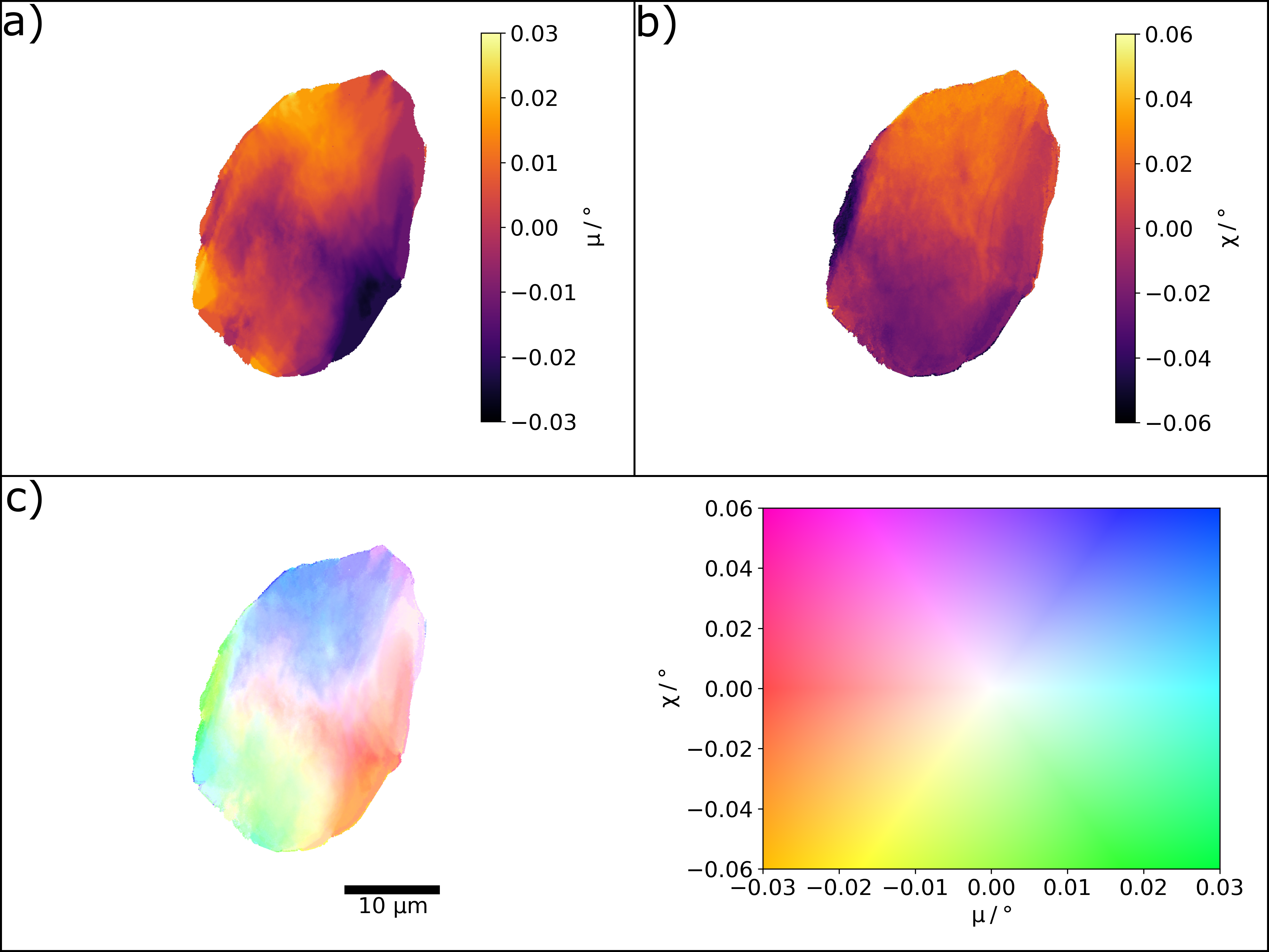}
    \caption{Grain 2 of the recrystallized iron sample imaged with the diamond lenses at \SI{33}{\keV}. a) Rocking angle $\mu$ at maximum intensity for each pixel. b) Rolling angle $\chi$ at maximum intensity for each pixel. c) Combined intragranular misorientation map referenced to the centre of the $\mu$ and $\chi$ rocking curves. The colormap and scale bar are indicated.}
    \label{rex_g2}
    \end{center}
\end{figure}

Figures~\ref{rex_g1} and~\ref{rex_g2} show mosaicity maps of two representative grains acquired using the diamond CRL objective at \SI{33}{\keV}, demonstrating the suitability of the optics for high-energy DFXM. The magnification was $M = 32.5$ with the $2\times$ far-field objective and the effective pixel size was \SI{200.0}{\nano\metre}.

The intragranular misorientation maps of the recrystallized iron grains show a narrow angular spread, as expected for fully recrystallized high-purity iron with a low dislocation density on the order of $10^{11}$–$10^{12}\,\mathrm{m^{-2}}$ \cite{sanna2026}. Over most of the grain interior, both the $\mu$- and $\chi$-dependent maps exhibit only weak orientation variations, indicating a low level of stored lattice curvature.

In contrast, increased intragranular misorientation is observed in the vicinity of the grain boundaries in both the $\mu$ and $\chi$ maps. These regions of enhanced angular spread reflect local lattice rotations induced by mechanical constraint and compatibility stresses arising from the interaction with neighbouring grains. The combined misorientation maps in Fig.~\ref{rex_g1} and Fig.~\ref{rex_g2} highlight this behaviour, with the largest deviations concentrated near the grain edges, while the grain interior remains comparatively uniform.

\subsubsection{Invar}
\begin{figure}
    \begin{center}
    \includegraphics[width=0.9\linewidth]{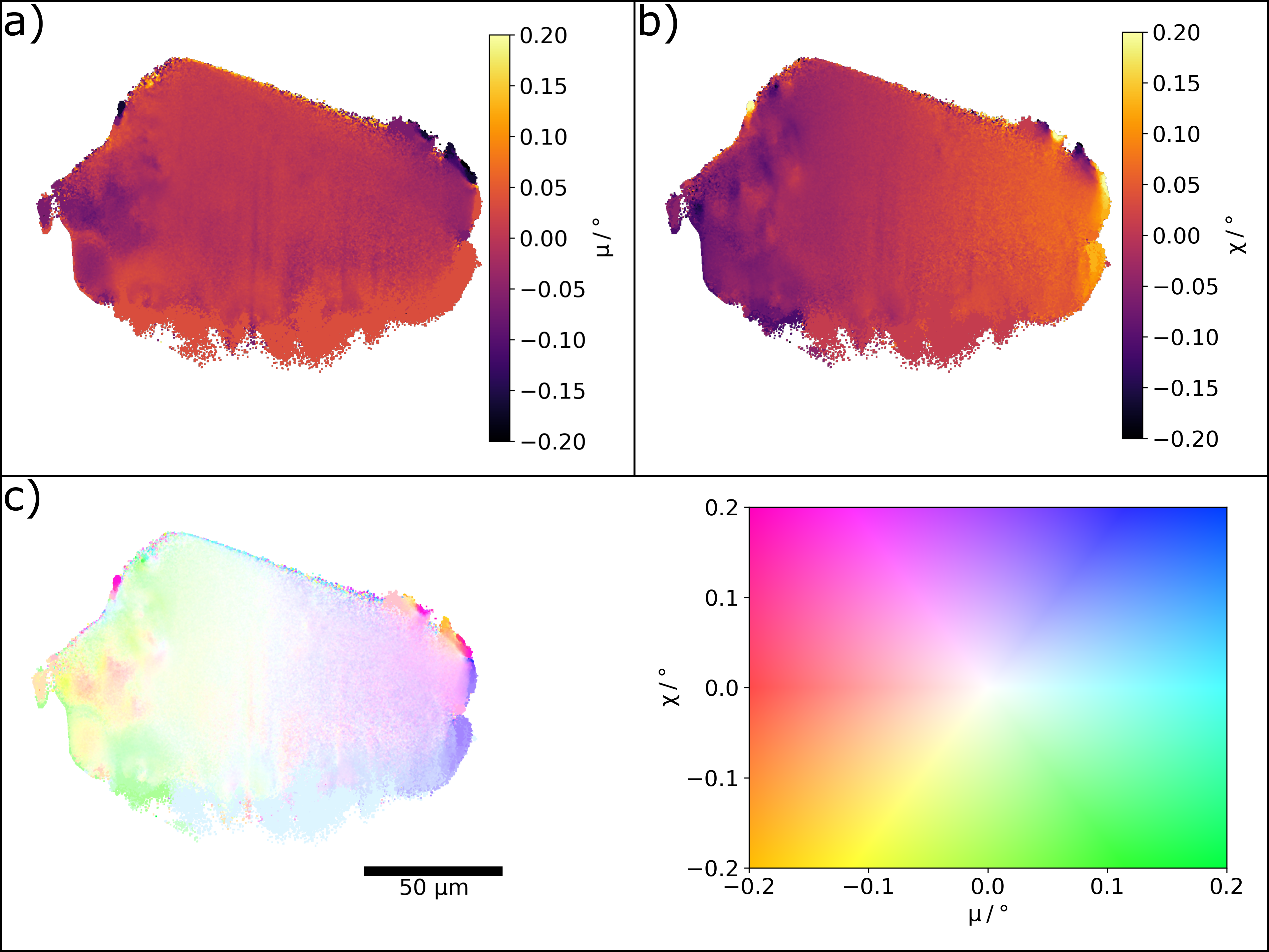}
    \caption{A grain within the Invar sample imaged with the diamond lenses at \SI{33}{\keV}. a) Rocking angle $\mu$ at maximum intensity for each pixel. b) Rolling angle $\chi$ at maximum intensity for each pixel. c) Combined intragranular misorientation map referenced to the centre of the $\mu$ and $\chi$ rocking curves. The colormap and scale bar are indicated.}
    \label{invar}
    \end{center}
\end{figure}

An Invar alloy sample was investigated using DFXM at a photon energy of \SI{33}{\keV} with the diamond CRL objective. Invar is an Fe–Ni alloy known for its low thermal expansion coefficient and features relatively high x-ray attenuation compared to pure iron, making it a representative test case for thicker and compositionally heavier materials. An Fe–36Ni Invar block (composition in wt.\%: 0.05 C, 0.6 Mn, 0.05 S, 0.40 Si, 38 Ni, 0.5 Co, 0.1 Al, 0.25 Cr, 0.10 Mg, 0.015 P, 0.10 Ti, 0.10 Zr, balance Fe) was heat treated at \SI{825}{\celsius} for \SI{1.5}{\hour}. Specimens were then extracted by electrical discharge machining and polished using up to a 1 \SI{1}{\micro \metre} diamond suspension. The material exhibits a microstructure consisting of equiaxed grains with sizes ranging from approximately \SIrange{20}{60}{\micro \metre}. The specimen has an approximately square cross-section with a thickness of about \SI{400}{\micro \metre}. A single grain within the equiaxed polycrystalline microstructure was selected for imaging, and intragranular mosaicity was mapped by scanning the rocking angle $\mu$ and the rolling angle $\chi$. Figure~\ref{invar} shows the resulting $\mu$- and $\chi$-dependent intensity maps together with the combined intragranular misorientation map. The magnification was $M = 162.5$ with the $10\times$ far-field objective and the effective pixel size was \SI{40.0}{\nano\metre}. In contrast to the recrystallized pure iron grain, the Invar grain exhibits a significantly larger intragranular orientation spread, on the order of $\sim0.4^\circ$, corresponding to approximately four times the angular spread observed in the recrystallized iron case. Despite this increased spread, the microstructure remains relatively well ordered, with only a limited number of subgrains and localized defect-related features visible in the misorientation maps.

Taken together, these examples demonstrate that diamond CRL objectives enable high-quality DFXM measurements through sample thicknesses approaching \SI{0.5}{mm} in iron- and nickel-based materials, which are not readily accessible at the photon energies typically employed with Be CRL optics. Operation at photon energies above \SI{30}{\keV} substantially reduces x-ray absorption and beam-induced heating compared to the \SI{17}{\keV} regime commonly used at ID03, thereby opening new experimental possibilities, including high-energy and pink-beam DFXM imaging \cite{Yildirim2025, Labella2025} modes for high $Z$ materials.



The results show that diamond CRLs are a viable alternative to common Be CRLs. At \SI{17}{\keV}, they exhibit a smaller FOV and slightly reduced contrast and therefore spatial resolution compared to the Be CRL currently used at ID03. In theory, Be is the superior material for basically all photon energies, but the difference in $\delta / \beta$ ratio gets smaller with increasing photon energy, so that the choice between diamond and beryllium is more ambiguous at \SI{33}{\keV}.At higher energies, the focal length of Be CRLs becomes prohibitively long as objective lenses. Instead, the diamond CRL is compared to an aluminium CRL at \SI{33}{\keV}, where it shows slightly better resolution and a larger FOV. None of the investigated lenses exhibits any relevant radial distortion.



\section{Conclusion}
Diamond CRLs for use as an objective in DFXM were characterized regarding their contrast with different resolution targets from \SIrange{100}{200}{\nano\metre} and compared to an existing Be CRL at \SI{17}{\keV} and to a Al CRL at \SI{33}{\keV}. At \SI{17}{\keV}, the Be CRL featured slightly better contrast (\num{0.0561 \pm 0.0018} compared to \num{0.0082 \pm 0.0008} vertical contrast at \SI{150}{\nano \metre}) and FOV. At \SI{33}{\keV}, the diamond CRL showed better contrast (\num{0.0331 \pm 0.0013} compared to \num{0.01578217 \pm 0.00000018} vertical contrast at \SI{150}{\nano \metre}, horizontal contrast similar) and FOV (\SI{125.11}{\micro \metre} to \SI{91.36}{\micro \metre}) than the Al CRL it was compared to. None of the CRLs showed significant radial distortion. For an application example and to showcase the new possibilities with an objective that works at higher photon energies such as \SI{33}{\keV}, a recrystallized iron sample and an Invar alloy sample were investigated with DFXM.\\
The use of these novel diamond CRL opens up new possibilities for DFXM. At \SI{33}{\keV}, thicker samples than before or samples containing heavy elements can now be investigated, which before exhibited too much attenuation. Given the limited availability of Be CRLs, diamond CRLs are a more than adequate alternative. Furthermore, diamond CRLs might also be applied as condensers for pink beam applications \cite{Yildirim2025,Labella2025}.

\vspace{5mm}\noindent\textbf{Acknowledgements}:  
Dhruv Anjaria (NSF grant nr.~2338346) is acknowledged for the Invar sample preparation.
C.Y.~acknowledges the financial support by the ERC Starting Grant ``D-REX'' nr.~101116911.
S.S.~acknowledges support by an ERC Advanced grant nr.~885022, by a Villum Investigator grant nr.~73771, by the Danish Agency for Science and Higher Education grant nr.~8144-0002.
This study has received financial support from the European Union:
AddMorePower (GA 101091621).

\referencelist[references]

\end{document}